\documentclass[letterpaper,twocolumn,preprintnumbers,secnumarabic,amsmath,amssymb,superscriptaddress]{revtex4}
\usepackage{amssymb,amsmath}
\usepackage{ulem} 
\usepackage{bm} 
\usepackage{booktabs} 
\usepackage{array}
\usepackage{latexsym}
\usepackage{graphicx}
\usepackage{color}
\usepackage{datetime}

\usepackage{verbatim}
\usepackage{chngpage} 

\usepackage{psfrag}

\usepackage{mciteplus}

\usepackage[colorlinks=true,      linkcolor=blue,      urlcolor=blue,      
            filecolor=blue,      citecolor=blue,       pdfstartview=FitH,     
						pdfpagemode=UseNone,      bookmarksopen=true]{hyperref}  
\usepackage[all]{hypcap}     

\begin{document}
\title{
Smooth horizonless geometries deep inside the black-hole regime  
}

 \author{Iosif Bena}
\affiliation{Institut de Physique Th\'eorique,  Universit\'e Paris Saclay,  CEA, CNRS, F-91191 Gif sur Yvette, France}

\author{Stefano Giusto} 
\affiliation{Dipartimento di Fisica ed Astronomia, Universit\`a di Padova \& INFN Sezione di Padova, Via Marzolo 8, 35131 Padova, Italy}

\author{Emil J. Martinec}
\affiliation{Enrico Fermi Inst. and Dept. of Physics, University of Chicago,  5640 S. Ellis Ave., Chicago, IL 60637-1433, USA}

\author{Rodolfo Russo} 
 \affiliation{Centre for Research in String Theory, School of Physics and Astronomy, Queen Mary University of London, Mile End Road, London, E1 4NS, United Kingdom }
 
\author{Masaki Shigemori} 
\affiliation{Yukawa Institute for Theoretical Physics, Kyoto University, Kitashirakawa-Oiwakecho, Sakyo-ku, Kyoto 606-8502 Japan}
 
\author{David Turton}
\affiliation{Institut de Physique Th\'eorique,  Universit\'e Paris Saclay,  CEA, CNRS, F-91191 Gif sur Yvette, France}

\author{Nicholas P. Warner}
\affiliation{Department of Physics and Astronomy and Department of Mathematics, University of Southern California, Los Angeles, CA 90089, USA\\ \vspace*{1mm}{\rm \textsf{ iosif.bena@cea.fr, stefano.giusto@pd.infn.it, ejmartin@uchicago.edu, r.russo@qmul.ac.uk, shige@yukawa.kyoto-u.ac.jp, david.turton@cea.fr, warner@usc.edu} }}

\begin{abstract}

We construct the first family of horizonless supergravity solutions that have the same mass, charges and angular momenta as general supersymmetric rotating D1-D5-P black holes in five dimensions. This family includes solutions with arbitrarily small angular momenta, deep within the regime of quantum numbers and couplings for which a large classical black hole exists.
These geometries are well-approximated by the black-hole solution, and in particular exhibit the same near-horizon throat. Deep in this throat, the black-hole singularity is resolved into a smooth cap.  We also identify the holographically-dual states in the ${\cal N}=(4,4)$ D1-D5 orbifold CFT\@.  Our solutions are among the states counted by the CFT elliptic genus, and provide examples of smooth microstate geometries within the ensemble of supersymmetric black-hole microstates.

\end{abstract}

\maketitle

\def\oneone{\rlap 1\mkern4mu{\rm l}}
\def\IC{\mathbb{C}}
\def\Neql#1{{\cal N}\!=\!{#1}}
\def\Nb{\overline{N}}
\def\Pb{\overline{P}}
\def\Qb{\overline{Q}}
\def\IR{\mathbb{R}}
\def\ZZ{\mathbb{Z}}
\def\flux{\Pi}
\def\cA{{\cal A}}
\def\cB{{\cal B}}
\def\cC{{\cal C}}
\def\cD{{\cal D}}
\def\cF{{\cal F}}
\def\cG{{\cal G}}
\def\cH{{\cal H}}
\def\cJ{{\cal J}}
\def\cL{{\cal L}}
\def\cM{{\cal M}}
\def\cN{{\cal N}}
\def\cP{{\cal P}}
\def\cQ{{\cal Q}}
\def\cR{{\cal R}}
\def\cS{{\cal S}}
\def\cU{{\cal U}}
\def\stratum{superstratum}
\def\strata{superstrata}
\def\nBPS#1{$\frac{1}{#1}$-BPS}
\def\NPW#1{{\bf \Red NPW:} {\Blue #1}}
\def\SG#1{{\bf \Red SG:} {\Blue #1}}
\newcommand{\emil}[1]{{\bf \Red EJM:} {\DarkBlue #1}}
\def\DT#1{{\bf \Red DT:} {\DarkGreen #1}}
\def\IB#1{{\bf \Red IB:} {\DB #1}}
\def\nBPS#1{$\frac{1}{#1}$-BPS}

\newcommand{\emilred}[1]{{\Red #1}}

\definecolor{cardinal}{rgb}{0.6,0,0}
\definecolor{darkgreen}{rgb}{0,0.45,0}
\definecolor{golden}{rgb}{0.92, 0.7, 0}
\definecolor{midnight}{rgb}{0, 0, 0.5}
\definecolor{darkblue}{rgb}{0.2, 0, 0.8}
\newcommand{\Red}{\color{red}}
\newcommand{\Blue}{\color{blue}}
\newcommand{\DarkGreen}{\color{darkgreen}}
\newcommand{\Cardinal}{\color{cardinal}}
\newcommand{\Golden}{\color{golden}}
\newcommand{\Midnight}{\color{midnight}}
\newcommand{\DarkBlue}{\color{darkblue}}
\def\DG{\DarkGreen}
\def\CR{\Cardinal}
\def\DB{\DarkBlue}

%

\newcommand{\bra}[1]{\langle#1|}
\newcommand{\ket}[1]{|#1 \rangle}

\def\({\left(}
\def\){\right)}
\def\[{\left[}
\def\]{\right]}

\def\ie{{i.e.}}
\def\eg{{e.g.}}
\def\cf{{c.f.}}
\def\etc{{etc}}

\def\cl{{\rm cl}}
\def\qu{{\rm qu}}
\def\ext{{\rm ext}}
\def\lpl{\ell_{\rm pl}}
\def\mpl{m_{\rm pl}}
\def\lstr{\ell_{\rm str}}
\def\gstr{g_{\rm s}}
\def\gym{g_{\rm \scriptscriptstyle YM}}

\def\sst#1{\scriptscriptstyle{#1}}

\def\bh{\mathrm{\sst{BH}}}
\def\btz{\mathrm{\sst{BTZ}}}

\def\phibtz{\varphi}
\def\lads{\ell_{\mathrm{\sst{AdS}}}}

\newcommand{\adstwo}{AdS$_2$ }
\newcommand{\adsthree}{AdS$_3$ }

\newcommand{\np}{\ensuremath{n_{\mathrm{\sst{P}}}}}
\newcommand{\qp}{\ensuremath{Q_{\mathrm{P}}}}

\def\half{\frac12}
\def\coeff#1#2{{\textstyle \frac{#1}{#2}}}
\def\hf{\coeff12}
\def\tr{{\rm Tr}}
\def\One{{\hbox{ 1\kern-.8mm l}}}

\def\barray{\begin{array}}
\def\earray{\end{array}}
\def\be{\begin{equation}}
\def\ee{\end{equation}}
\def\bea{\begin{eqnarray}}
\def\eea{\end{eqnarray}}
\def\bal{\begin{align}}
\def\eal{\end{align}}
\def\nn{\nonumber}

\section{Introduction}
\label{Sect:Introduction}
\vspace{-1mm}

The black-hole information paradox reveals a profound conflict between Quantum Mechanics and General Relativity~\cite{Hawking:1976ra}.  Quantum mechanically, a black hole has an entropy given by the horizon area in Planck units, while in General Relativity the black hole is unique for a given mass, charge and angular momentum.  Unitarity is violated because the enormous black-hole entropy is not visible at the black-hole horizon and so the information about the black-hole state cannot be encoded in the Hawking radiation.  Thus, unitarity can only be preserved if there is new physics at the scale of the horizon~\cite{Mathur:2009hf}. However, constructing structure at the scale of the horizon is no easy task: The horizon is a null surface, and any classical matter or wave that can carry information will either fall in or dilute very fast. 

One of the great successes of string theory has been a precise
accounting of the entropy of certain black holes~\cite{Strominger:1996sh}, 
and the identification of the microstates that
give rise to this entropy, albeit in a regime of coupling where the classical
black-hole solution is not valid.  However, this is not enough to solve
the information paradox. In order to create the required structure at the horizon, all the typical microstates of the black hole must become horizon-sized bound states 
that have the same mass and conserved charges as the black hole, and that exist 
in the same regime of parameters in which the classical black-hole solution is valid. 
Furthermore, microstates that are describable in supergravity should be horizonless.

For supersymmetric black holes it has been possible to construct  large classes of supergravity solutions corresponding to such horizonless bound states, and these are known as ``fuzzball'' or microstate geometries~\cite{Mathur:2005zp,Bena:2007kg}. These solutions correspond to {\it some\/} of the microstates of the black hole, but have limitations, as we now discuss.

The microstate geometries constructed in~\cite{Bena:2005va,Berglund:2005vb,Bena:2006kb,Bena:2007qc}, although carrying
the same charges and angular momenta as a large black hole, 
have the following issues:
(i) In all examples, these solutions carry an angular momentum that is a large fraction of the maximally allowed value for the black hole;
(ii) Their CFT dual is not known and so their role in the ensemble of black-hole microstates remains unclear; 
(iii) It is not clear whether these configurations are generic and represent typical microstates of a black hole~\cite{deBoer:2009un}, nor whether the states of the black hole will continue to be described by such geometries when the black hole becomes non-extremal. 

Another class of microstate geometries relevant for large supersymmetric black holes in five dimensions is discussed in~\cite{Lunin:2012gp,Giusto:2013bda,Bena:2015bea,Bena:2016agb}.  While these solutions have known CFT duals, they also carry macroscopic five-dimensional angular momenta $j, \tilde{j}$. 

The purpose of this Letter is to simultaneously resolve 
the first two issues described above by (i) constructing the first microstate geometries of rotating, supersymmetric D1-D5-P (BMPV) black holes in string theory~\cite{Breckenridge:1996is} in which the angular momenta take arbitrary finite values, in particular including arbitrarily small values; and by (ii) identifying the dual CFT states. In doing so we also demonstrate, via an explicit example, that adding momentum to a two-charge solution describing a microstate of a string-size black hole can result in a large-scale, low-curvature supergravity solution.

\section{Black-hole microstate geometries}
\vspace{-1mm}

We work in type IIB string theory on $\mathbb{R}^{4,1}\times$S$^1\times \cM$, where $\cM$ is  T$^4$ or $K3$. We take the size of $\cM$ to be microscopic, and that of S$^1$ to be macroscopic. The S$^1$ is parameterized by the coordinate $y$. We wrap $n_1$ D1-branes on the S$^1$ and $n_5$ D5-branes on S$^1\times \cM$, and consider momentum charge, P, along the $y$ direction. We work in the low-energy, six-dimensional supergravity theory obtained by reduction on $\cM$.

The near-horizon geometry of a six-dimensional rotating, supersymmetric black string with the foregoing charges is S$^3$ fibered over the extremal BTZ black hole~\cite{BTZ}, whose metric is:
\be
\label{dsBTZ}
ds^2_{\btz} =\lads^2\Bigl[\rho^2(-dt^2+dy^2) + \frac{d\rho^2}{\rho^2} + \rho_*^2(dt+dy)^2 \Bigr] .
\ee
This metric is locally \adsthree and it asymptotes to the standard \adsthree  form for $\rho\gg \rho_{\ast}$. It can be written as a circle of radius $\rho_{\ast}$ fibered over \adstwo  in the near-horizon region $\rho\ll \rho_{\ast}$ (see, for example,~\cite{Strominger:1998yg}). Dimensional reduction on this circle yields the \adstwo of the near-horizon BMPV solution. Following the usual abuse of terminology, we will refer to this region as the \adstwo throat.

The BTZ parameters are related to the supergravity D1, D5, and P charges $Q_{1,5,P}$ and the radial coordinate $r$ (to be used later) via $\rho=r/\sqrt{Q_1Q_5}$ and $\lads^2=\sqrt{Q_1Q_5}$. The horizon radius, $\rho_\ast$, of the BTZ solution \eqref{dsBTZ} determines the onset of the \adstwo throat (and thus the radius of the fibered $S^1$) and is given by $\rho_\ast^2 = \qp/(Q_1Q_5)$. This value is determined by a competition between the momentum charge  that exerts pressure on the geometry, and the D1 and D5 charges that exert tension. 

Typical black-hole microstates should be very well-approximated by the black-hole solution until very close to the horizon. This requires a long, large, BTZ-like \adstwo throat.
To obtain such a throat, prior work has used bubbling solutions with multiple Gibbons-Hawking (GH) centers%
~\cite{Bena:2005va,Berglund:2005vb}; 
the moduli space of these solutions  includes ``scaling'' regions
\cite{Denef:2002ru,Bena:2006kb,Bena:2007qc}
in which the GH centers approach each other arbitrarily closely, whereupon the solution develops an arbitrarily long \adstwo throat.
It has been argued that quantum effects set an upper bound on the depth of such throats~\cite{Bena:2007qc,deBoer:2008zn}, and a corresponding lower bound on the energy gap, which matches the lowest energy excitations of the (typical sector of the) dual CFT\@.
This suggests that microstate geometries are capable of sampling typical sectors of the dual CFT.

Unfortunately, all the previously-known scaling microstate geometries involve at least three GH centers, whose dual CFT states are currently unknown.  The holographic dictionary between supergravity solutions and CFT states has been constructed only for two-centered solutions~\cite{Giusto:2012yz}; we therefore construct new black-hole microstate solutions by adding momentum excitations to a certain two-charge seed solution.  We do this using ``superstratum'' technology~\cite{Bena:2014qxa,Bena:2015bea,Bena:2016agb} to introduce deformations, with specific angular dependence, so as to modify the momentum and the angular momenta of the solution. 

A particular sub-class of our deformations has the effect of reducing the angular momenta of the two-charge seed solution, while introducing no additional angular momentum. These deformations therefore allow us to obtain solutions that have arbitrarily small angular momenta and describe microstates of the non-rotating D1-D5-P (Strominger-Vafa) black hole.  The solutions have an \adstwo throat, which becomes longer and longer as the angular momenta $j,\tilde{j} \to 0$, thus classically approximating the non-rotating black hole to arbitrary precision.

\vspace{-2mm}
\section{The new class of solutions}
\label{Sect:Solution}
\vspace{-1mm}

The metric, axion and dilaton of our \nBPS{4} solutions are determined by four functions, $Z_1, Z_2, Z_4,\cF$ and two  vector fields $\beta, \omega$ \cite{Giusto:2013rxa}:
\begin{equation}
ds_6^2 =   -\frac{2}{\sqrt{\cP}} \, (dv+\beta) \big(du +  \omega + \tfrac{1}{2}\, \mathcal{F} \, (dv+\beta)\big) 
~+~  \sqrt{\cP} \, ds_4^2\,,  \label{sixmet}
\end{equation}
where $ds_4^2$ is the flat metric on $\IR^4$ written in spherical bipolar coordinates,
\begin{equation}
ds_4^2 = \frac{\Sigma\,  dr^2}{r^2 + a^2} +  \Sigma\, d \theta^2  +  (r^2 + a^2) \sin^2 \theta \, d\phi^2 \\
+ r^2  \cos^2 \theta \, d\psi^2  ,
\label{bipolmet}
\end{equation}
with $0\leq \theta \leq \pi/2$ and $0\leq \phi,\psi<2 \pi$. The coordinates $u$ and $v$ are light-cone variables related to the asymptotic time $t$ and the S$^1$ coordinate $y$ via:
\begin{equation}
u \equiv (t-y)/\sqrt{2} \,,  \quad v \equiv  \, (t+y)/\sqrt{2} \,, \quad y \cong  y +  2 \pi R_y  \,. 
\label{tyuv}
\end{equation}
The functions  $\Sigma$ and $\cP$ are defined by:
\begin{equation}
\Sigma \equiv  r^2 + a^2 \cos^2 \theta  \,, \quad \cP   \equiv    Z_1 \, Z_2  -  Z_4^2    \,,
\label{PSigDefn}
\end{equation}
and the dilaton and axion are given by:  
\begin{equation}
 e^{2\Phi} =  Z_1^2\,\cP^{-1}\,,  \quad     C_0 =Z_4 Z_1^{-1}   \,.
\label{DilAx}
\end{equation}
The tensor gauge fields are also related to these functions but we will not discuss their explicit form here. 

We consider solutions that have a simple $v$-fibration: 
\begin{equation} 
\beta \, =\, 2^{-1/2}\, a^2 R_y \, \Sigma^{-1} \, ( \sin^2 \theta \, d\phi-   \cos^2 \theta \, d\psi ) \,.
\label{betaform1}
\end{equation} 

We begin with the background of a maximally-rotating D1-D5 supertube~\cite{Balasubramanian:2000rt,Maldacena:2000dr} and add deformations that  depend upon the  angles $(v, \phi,\psi)$  via the phase dependence:
\begin{equation} 
 \hat{v}_{k,m,n} \equiv \sqrt{2}\, R_y^{-1}\, (m+n)  \,v+ (k-m)\phi - m\psi \,, \label{phase1}
\end{equation} 
where $k\in\mathbb{Z}_{>0}$ and $m,n\in\mathbb{Z}_{\ge 0}$. These
fluctuations modify the angular momenta $j,\tilde{j}$ and the momentum
number $\np=p_y R_y$ with $p_y$  the momentum along the $y$ circle.  
In order to obtain smooth solutions whose holographic duals we can identify, 
we add a fluctuating mode with strength $b_{k,m,n}$
using the ``coiffuring'' technique of
\cite{Giusto:2013bda,Bena:2014rea,Bena:2015bea,Bena:2016agb}:
\begin{align} \label{Z1}
Z_1 & =  \frac{Q_1}{\Sigma} + \frac{R_y^2}{2 Q_5} b_{k,m,n}^2  \frac{\Delta_{2k,2m,2n}}{\Sigma} \cos \hat{v}_{2k,2m,2n} \,, \\ \label{Z2Z4}
Z_2  &= \frac{Q_5}{\Sigma} \,, \quad 
Z_4  =  b_{k,m,n} R_y\frac{\Delta_{k,m,n}}{\Sigma} \cos \hat{v}_{k,m,n} \,,
\end{align} 
where
\begin{equation} 
\Delta_{k,m,n} \equiv  a^k \, r^n (r^2+a^2)^{-(k+n)/2} \cos^{m}\theta \, \sin^{k-m}\theta \,.  \label{Deltadefn}
\end{equation} 
This coiffuring ensures that, while the tensor fields  depend on $ \hat{v}_{k,m,n} $, the metric does not. The remaining parts of the solution are given by 
\begin{equation}
\cF  = b^2_{k,m,n}\, \cF_{k,m,n}  \,, \quad \omega = \omega_0 +   b^2_{k,m,n} \,\omega_{k,m,n} \,,
\end{equation} 
where $\omega_0$ is the value that $\omega$ takes in the undeformed supertube solution:
\begin{equation} 
\omega_0 \,\equiv\, 2^{-1/2}\, a^2 \, R_y\, \Sigma^{-1}\, ( \sin^2 \theta \, d\phi +   \cos^2 \theta \, d\psi ) \,.
\label{omegaring}
\end{equation} 
The general expressions for $\cF_{k,m,n}$ and $\omega_{k,m,n}$ are given in Appendix~\ref{Sect:Appendix} and we leave the expressions of the tensor gauge fields to a subsequent publication.

Regularity and absence of closed timelike curves (CTCs) requires
\begin{equation} 
\label{strandbudget1}
Q_1Q_5/R_y^2=a^2 + b^2/2 \,, \qquad b^2 = x_{k,m,n} \,  b_{k,m,n}^2\,,
\end{equation} 
with $x_{k,m,n}^{-1}\equiv {k \choose m} {k+n-1 \choose n}.
$
The conserved charges of the solution are 
\begin{equation}
  \label{eq:jljr}
  j =  \frac{\mathcal{N}}{2} \left({a^2} + \frac{m}{k} \, b^2\right), \hspace*{.5cm} \tilde{j} = \frac{\mathcal{N}}{2} a^2,  \hspace*{.5cm}  \np  = \frac{\mathcal{N}}{2} \frac{m+n}{k} \, b^2\;
\end{equation}
where $ \mathcal{N} \equiv n_1 n_5 R_y^2 / (Q_1 Q_5)$, with $n_1, n_5$ the numbers of D1 and D5 branes. 

Rotating D1-D5-P black holes with regular horizons exist when $n_1 n_5 \np-j^2>0$ and this cosmic censorship bound  defines the ``black-hole regime'' for these parameters. Our solutions lie within this bound for
\begin{equation}
  \label{eq:bhbound}
  \frac{b^2}{a^2}> \frac{k}{n+\sqrt{(k-m+n)(m+n)}}\,.
\end{equation}
Hence, in this regime of parameters, these solutions correspond to horizonless microstates of large-horizon-area BMPV black holes.  They span the whole range of angular momenta that these black holes can have.  This is a dramatic improvement over the earlier solutions \cite{Bena:2006kb,Bena:2007qc}, which only have $j \gtrsim 0.88 \sqrt{n_1 n_5 \np}$. The solutions with $m=0$ are also remarkable because, as $a \rightarrow 0$, they give the first family of microstate geometries of the non-rotating D1-D5-P black hole. An explicit example  (with $k=1, m=0$ and general $n$) is given by:
\begin{equation}
\begin{aligned}
\label{sol10n}
\cF_{1,0,n} & = - a^{-2} \left(1 - r^{2n}(r^2+a^2)^{-n}\right) \\ 
\omega_{1,0,n} & =    2^{-1/2}\, R_y\, \Sigma^{-1}
 \left(1 - r^{2n}(r^2+a^2)^{-n}\right) \, \sin^2\theta\, d\phi\,. 
\end{aligned}
\end{equation}
One can easily show that the corresponding metrics are regular and have no CTCs. For our more general class of solutions, this proof becomes increasingly complicated, however our construction explicitly removes CTCs in the most dangerous regions (near $r=0$ and  $\theta=0$ or $\pi/2$), and there is little reason to expect problems elsewhere. 

\section{The dual CFT states}
\label{Sect:CFT}
\vspace{-1mm}

Our geometries are asymptotically AdS$_3\times$S$^3$ and correspond holographically  to
 $1/2$-BPS states in a $(4,4)$  two-dimensional CFT with central charge $c=6 n_1 n_5 \equiv 6N$. 
 Since these states are supersymmetric, they should have a simple description
 at the locus in moduli space at which the CFT is realized
 as the symmetric orbifold ${\cal M}^N/S_N$. 
 
The untwisted sector of this theory consists of $N$ copies of the CFT with target space $\mathcal{M}$. The theory also contains twisted sectors, in which the elementary fields have non-trivial periodicities connecting different CFT copies: When $k$ copies are cyclically permuted by the boundary conditions, we call the corresponding state a ``strand of winding $k$''.  Following the conventions of~\cite{Giusto:2015dfa}, we denote by $|\!+\!+\rangle_1$ a strand of length $1$ in the RR ground state that has $j=\tilde{j}=1/2$.
To describe our states we will also need the twisted-sector RR ground state, $|00\rangle_k$, which has winding $k$ and is a scalar under all symmetries. 

In our solutions, the momenta are carried by excitations of the $|00\rangle_k$ strands. These excitations can be described by $(4,4)$ 
superconformal algebras with central charge $6k$ living on each strand.  Denoting the Virasoro generators as $L_{n}$ and the R-symmetry
 $SU(2)$ generators as $J^{i}_n$, we excite the $|00\rangle_k$ strands with two
mutually-commuting, momentum-carrying perturbations: $J^{+}_{-1}= (J^1_{-1} + i J^2_{-1})$
 and $(L_{-1} - J^3_{-1})$. 

The charges of our solutions~(\ref{eq:jljr})
 support the identification of the dual CFT states with coherent superpositions of states of the form:
 \begin{equation}
  \label{eq:o3cstate}
(|\!+\!+\rangle_1)^{N_1 }
\biggl(\frac{(J^+_{-1})^{m}}{m!} \frac{(L_{-1}- J^3_{-1})^n}{n!} |00\rangle_k\biggr)^{N_{k,m,n} }\,,
\end{equation}
for all values of $N_1$ such that $N_1+  k N_{k,m,n} = N$. 

To find the exact coefficients of this superposition of states, one can straightforwardly generalize the derivation of~\cite{Giusto:2015dfa} to states with $n>0$. These coefficients are thereby determined
in terms of the supergravity parameters $a$ and $b_{k,m,n}$. 

From this calculation one finds that the average numbers of $|\!+\!+\rangle_1$ and $|00\rangle_k$ strands are given by $\mathcal{N}a^2$ and  $\mathcal{N}b^2/(2k)$ respectively, from which the strand quantum numbers immediately yield the supergravity momentum and angular momenta in~(\ref{eq:jljr}).  
It is also possible to compute 3-point correlators between our heavy states and BPS states of low conformal dimension. These correlators depend not only on the 
 average numbers of strands but also on the spread of the coherent superposition, providing an even more stringent check of the identification between the CFT states and the supergravity solutions.

\section{The structure of the metric}
\label{Sect:Metric}
\vspace{-1mm}

In the AdS/CFT limit, one takes $R_y$ to be the largest scale in the problem, implying that $\qp \ll \sqrt{Q_1Q_5}$.  
We further focus on the regime  $a^2 \ll \qp = (m+n)b^2/(2k)$, in which the structure of the cap lies deep
inside the \adstwo region discussed above. 

In the $a \to 0 $ limit, the \adstwo  throat tends to infinite depth and our solutions tend to the BMPV solution. Furthermore, when $m=0$, the angular momentum vanishes and the solutions tend to that of the non-rotating D1-D5-P (Strominger-Vafa) black hole.

\begin{figure}
\centerline{\includegraphics[width=3.0in]{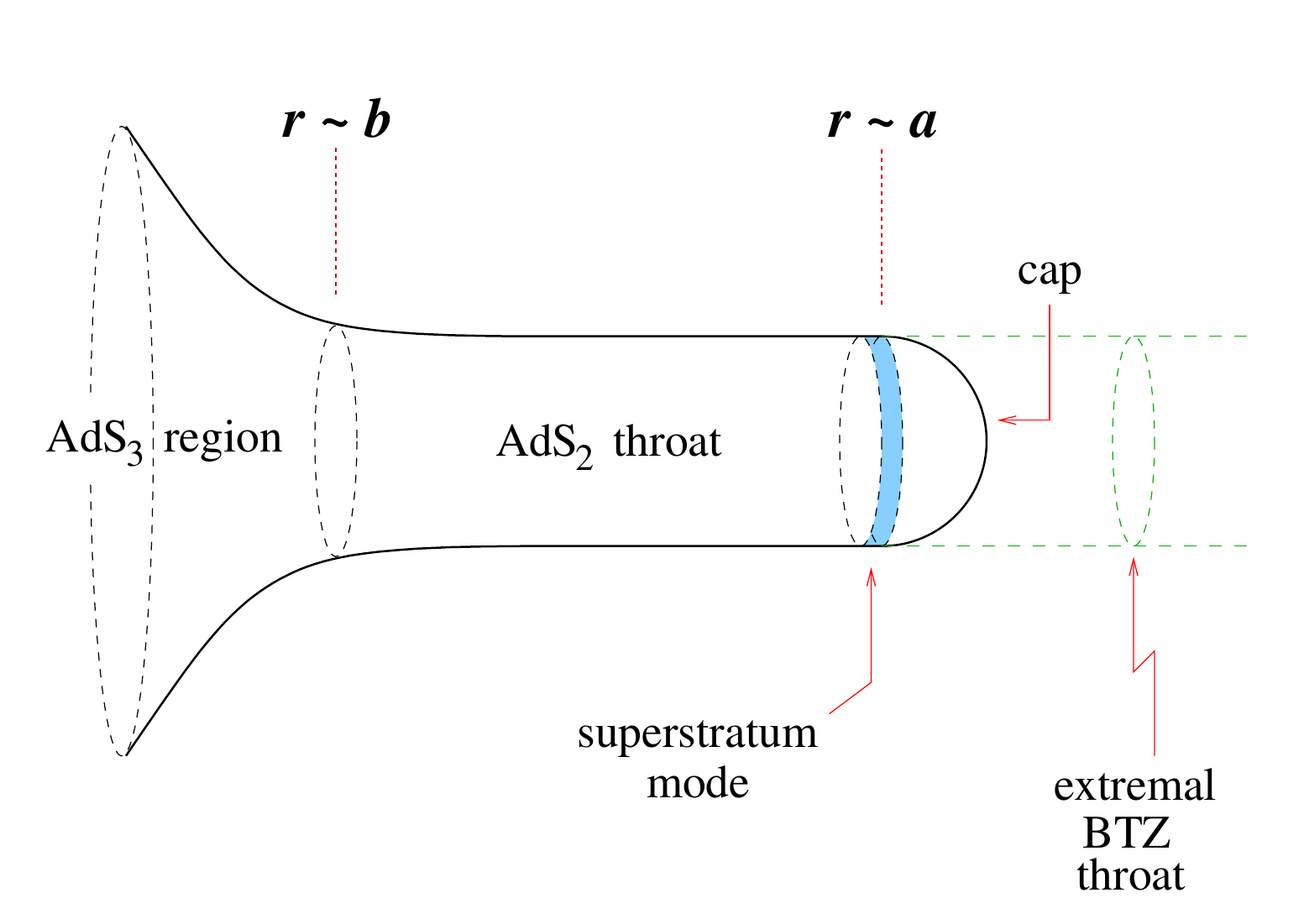}}
\setlength{\unitlength}{0.1\columnwidth}
\caption{\it 
Sketch of the superstratum spatial geometry in the $r$-$y$ plane, compared to the extremal BTZ geometry.}
\label{fig:ExtremalBTZ}
\end{figure}

If, instead, we keep $a^2 \ll \qp $ small but finite, then the leading terms in the metric for $r\gg a$ are those of the corresponding black hole. The \adstwo  throat extends in the radial direction for a proper length of order $\lads\log{(\qp/a^2)}$, and the geometry caps off smoothly in the region $r\ll a$, as shown in Fig.\;\ref{fig:ExtremalBTZ}.
In string units, the proper length of the $y$ circle in the \adstwo throat is of order $(g_s \np)^{1/2}/N^{1/4}$ when the volume of the compact space ${\cal M}$ is of order one. 
Thus one can easily arrange that the proper length of the $y$ circle in the \adstwo  region is large in string units, whereupon the supergravity approximation is valid.

%
The momentum charge is carried by a superstratum deformation (supergravity wave) concentrated deep inside the \adstwo  region. The wave profile is determined by the functions $\Delta_{k,m,n}$~\footnote{To see this, note that the $\Delta_{k,m,n}$ are peaked around $r_{\rm max} \sim a\sqrt{n/k}$ (and around a band of latitude on $S^3$), so for instance for $n\sim k$ and $n$, $k$ large, the wave profile is sharply peaked near $r\sim a$ (the blue band in Fig.\;\ref{fig:ExtremalBTZ}).}.
Inside the support of the wave, the momentum density that stabilized the size of the $y$-circle quickly dilutes, and the circle starts to shrink until one gets to $r=0$, where the coiffuring relations guarantee that the geometry caps off smoothly.

Our solutions therefore provide examples, with arbitrary finite angular momenta, of how the horizon of a D1-D5-P black hole can be replaced by a smooth cap. The solutions only differ significantly from the corresponding black hole metric near the cap; the difference is suppressed in the AdS$_2$ throat and further out into the asymptotic \adsthree region. For example, in the $k=1$, $m=0$, general $n$ solution, the leading corrections to the corresponding black hole metric have magnitude (in a local orthonormal frame) of order $\sqrt{n}\,a^2/r^2$ in the \adstwo throat, and of order $a^2/r^2$ in the asymptotic \adsthree region. 

When $a$ is exactly zero, from the dictionary \eqref{eq:jljr} one can see that for any value of $j$ and $\np$ there exists a one-parameter family of CFT states that should correspond to a bulk solution with $a=0$. Since this solution is exactly the classical black-hole solution with an event horizon, one might naively conclude that certain pure CFT states have a bulk dual with an event horizon, which would contradict the intuition expressed in the Introduction. However, several hints indicate that the strong-coupling description of these particular states (and also of two-charge states of the form  $(|00\rangle_k)^{\frac{N}{k}}$) requires ingredients beyond supergravity.  For example, the supergravity approximation to the  sequence of dualities used to derive the geometry~\cite{Lunin:2001fv} is not valid in these instances.  Moreover, in the D1-D5 CFT, this class of states can be distinguished from the thermal ensemble only by the VEVs of non-chiral primary operators. 

\vspace{-1mm}
\section{Discussion}
\label{Sect:Disc}
\vspace{-1mm}

In this Letter, we have constructed a new family of black hole microstate geometries 
that solve the ten-year-old problem of lowering the angular momentum $j$ arbitrarily 
below the cosmic censorship bound, and we have identified the dual CFT states. 
Our results demonstrate how adding momentum can transform a two-charge solution 
describing a microstate of a string-size black hole into a smooth low-curvature solution 
with a long \adstwo throat.
We are confident that all the solutions one can build by generalizing
the present work to include more general fluctuations
will continue to share these properties.
The generic black-hole microstate differs from the states we have
constructed in the distribution and type of momentum carriers -- our
solutions correspond in the CFT to using a very limited set of generators of the
chiral algebra (see (\ref{eq:o3cstate})) to carry the momentum.  It is a
very interesting question to ask how closely one can approach the
generic state using our techniques.

\vspace{5mm}
\noindent
{\bf Acknowledgments}.  
%
We thank Samir Mathur for discussions.
The work of IB and DT was supported by the John Templeton Foundation Grant 48222. 
The work of EJM was supported in part by DOE grant DE-SC0009924. 
The work of SG was supported in part by the Padua University Project
CPDA144437. 
The work of RR was partially supported by the STFC Consolidated Grant ST/L000415/1 \textit{
``String theory, gauge theory \& duality''}.
The work of MS was supported in part by 
JSPS KAKENHI Grant Number 16H03979.
The work of DT was supported in part by a CEA Enhanced Eurotalents Fellowship. 
The work of NPW was supported in part by the DOE grant
DE-SC0011687.
SG, EM, RR, MS and NPW are very grateful to the IPhT, CEA-Saclay for
hospitality while a substantial part of this work was done.

\newpage
\appendix
\begin{widetext}
%
\section{Details of the general solution}
\label{Sect:Appendix}
\vspace{-1mm}
The form of $\cF_{k,m,n}$ and $\omega_{k,m,n}$ for general $k,m,n$ is
\begin{equation} \label{cF}
\cF_{k,m,n} = 4\biggl[\frac{m^2 (k+n)^2}{k^2}\,F_{2k,2m,2n}+\frac{n^2 (k-m)^2}{k^2}\,F_{2k,2m+2,2n-2}\biggr],~~ \omega_{k,m,n}= \mu_{k,m,n}\,(d\psi+d\phi)+\zeta_{k,m,n}\,(d\psi-d\phi)\,,
\end{equation} 
\vspace{-2mm}
\begin{equation} \label{mu}
\mu_{k,m,n}= \frac{R_y}{\sqrt{2}}\,\biggl[ 
\frac{(k-m)^2(k+n)^2}{k^2} F_{2k,2m+2,2n}
+\frac{m^2 n^2}{k^2} F_{2k,2m,2n-2}
-\frac{r^2+a^2\,\sin^2\theta}{4\,\Sigma}\mathcal{F}_{k,m,n}
-\frac{\Delta_{2k,2m,2n}}{4\,\Sigma}
+\frac{x_{k,m,n}}{4\,\Sigma}
\biggr]\,,
\end{equation} 
where
\vspace{-2mm}
\begin{equation} 
F_{2k,2m,2n}\,=\,-\!\sum^{j_1+j_2+j_3\le k+n-1}_{j_1,j_2,j_3=0}\!\!{j_1+j_2+j_3 \choose j_1,j_2,j_3}\frac{{k+n-j_1-j_2-j_3-1 \choose k-m-j_1,m-j_2-1,n-j_3}^2}{{k+n-1 \choose k-m,m-1,n}^2}\,
\frac{\Delta_{2(k-j_1-j_2-1),2(m-j_2-1),2(n-j_3)}}{4(k+n)^2(r^2+a^2)}\,,
\end{equation} 
%
\end{widetext}
and where
\vspace{-1mm}
\begin{equation} 
{j_1+j_2+j_3 \choose j_1,j_2,j_3}\equiv \frac{(j_1+j_2+j_3)!}{j_1! j_2! j_3!}\,.
\end{equation} 
It should be understood that in $\mathcal{F}_{k,m,n}$ and $\mu_{k,m,n}$, when the coefficient of an 
$F$ function is zero, the term is zero.

The expression for  $\zeta_{k,m,n}$  can be obtained from  $\mu_{k,m,n}$  by quadrature using the BPS equations for $\omega$, which now reduce to an integrable system of differential equations, as was the case for the $n=0$ solutions studied in~\cite{Bena:2015bea}. 

For regularity, $\mu_{k,m,n}$ must vanish at $r=0, \theta =0$; this fixes $x_{k,m,n}$ to the value given below Eq.\;(\ref{strandbudget1}).

One might worry that the warp factor $Z_1$ could become negative and render the solution singular
if the amplitude of the fluctuations becomes too large. However, the minimal value of $Z_1$ occurs when $\cos\hat{v}_{2k,2m,2n}=-1$. Then the regularity conditions in Eq.\;(\ref{strandbudget1}) and the identity
\begin{equation} \nonumber
\frac{\Delta_{2k,2m,2n}}{x_{k,m,n}} \le\!\! \sum_{m,n=0\atop k=1}\delta_{k+n-1,p}\, \frac{\Delta_{2k,2m,2n}}{x_{k,m,n}}  ~=~  \frac{a^2 }{(r^2 +a^2)} \le 1\,
 \label{combinatoric1} 
\end{equation} 
ensure that $b^2_{k,m,n} \Delta_{2k,2m,2n} < b^2$ and hence $Z_1>0$.


\end{document}